\documentclass[conference]{IEEEtran}
\usepackage{cite}

\ifCLASSINFOpdf
  \usepackage[pdftex]{graphicx}
  \graphicspath{{../pdf/}{../jpeg/}}
   \DeclareGraphicsExtensions{.pdf,.jpeg,.png}
\else
\fi
\usepackage[cmex10]{amsmath}
\usepackage{xfrac}
\usepackage{array}

\usepackage{mdwmath}
\usepackage{mathptmx}
\usepackage{amsfonts}
\usepackage{amssymb}
\usepackage{amsbsy}
\usepackage{amsthm}

\newtheorem{lemma}{Lemma}
\usepackage{footnote}
\makesavenoteenv{tabular}
\usepackage{subcaption}
\usepackage{caption}
\usepackage{longtable}

\begin{document}
%
\title{A Generalized Write Channel Model for Bit-Patterned Media Recording}




\author{\IEEEauthorblockN{Sima Naseri \IEEEauthorrefmark{1}\IEEEauthorrefmark{2},
Somaie Yazdani \IEEEauthorrefmark{1}\IEEEauthorrefmark{2},
Behrooz Razeghi \IEEEauthorrefmark{1},
Ghosheh Abed Hodtani \IEEEauthorrefmark{1}}
\IEEEauthorrefmark{1} Department of Electrical Engineering, Faculty of Engineering,
Ferdowsi University of Mashhad,
Mashhad, Iran\\
\IEEEauthorrefmark{2} Communication and Computer Research Center, Ferdowsi University of Mashhad,
Mashhad, Iran\\
E-mail: naseri.sima1990@gmail.com, yazdani.somaie@gmail.com, behrooz.razeghi.r@ieee.org, ghodtani@gmail.com
 }



\maketitle

\begin{abstract}
In this paper, we propose a generalized write channel model for bit-patterned media recording by considering all sources of errors causing some extra disturbances during write process, in addition to data dependent write synchronization errors. We investigate information-theoretic bounds for this new model according to various input distributions and also compare it numerically to the last proposed model.

\end{abstract}


%
\IEEEpeerreviewmaketitle
\vspace{-4pt}
\section{Introduction}
Magnetic recording channels have been one of the most important media for data recording due to their capability of high density recording. There are various ways for magnetic data storage in such channels, particularly two dimensional magnetic recording (TDMR) \cite{wood2009feasibility, ckrishan2009readchannel, chan2010models}, heat assisted magnetic recording (HAMR) \cite{kryder2008heat}, microwave assisted magnetic recording (MAMR) \cite{zhu2008microwave} and bit-patterned media recording (BPMR) \cite{terris2006cooperative}.  BPMR, as the recent method for this aim, offers some extra merits to this process such as high density recording as a result of its improved thermal stability \cite{richer2006density}. However, some errors exclusive to these media are always a matter of concern particularly in read back process \cite{nutter2008understanding}. There are a number of different factors causing write errors, especially in bit-patterned media recording. One of these popular factors is the problem of write synchronization errors \cite{richer2006density}, which is detrimental during writing process since it necessitates the write pulses to be synchronized with the discrete and predetermined position of bits \cite{hu2007limits}.

 In \cite{zhang2010failure}, write failure analysis was presented based on two main sources of errors occurring in bit-patterned media, and was proved that written-in errors could stem from exceeding shift of writing window due to increasing switching field of the grain, and the poor head field which is lower than the switching field. While, the former results in a data-dependent error, relating to the intended recorded bit, the latter contributes to a random error without any dependency on the last written data. Therefore, it is expected to confront an extra failure error during writing process, which does not just depend on the last written bit as in the channel proposed in \cite{iyengar2011writechannel}, based on both substitution-like and insertion-deletion errors. In \cite{hu2007limits} and \cite{iyengar2009lpdc}, the binary symmetric channel was proposed as the write channel, where inserted random bit is considered as an error in such media.
\vspace{-2pt}

\textbf{Our work}:
In this paper, we introduce a more general write channel model by considering a binary noise and attempt to find information-theoretic bounds for this channel according to the rule of writing.

The characteristics of magnetic materials necessitate some special features to be considered. According to the high coercivity of some magnetic materials, a strong magnetic field is needed to record data on the disk. When data is recorded, the amount of magnetization that remains after magnetic field is removed would be important to ensure the stability of recording data on the disk.

In pre-patterned media, there is recorded data, before writing the desired bits on the media. The insufficient head field is one of the most important factors involved in write errors \cite{muraoka2011statistical}. As a matter of fact, the current written bit's magnetic field must be large enough to dominate the reversal field of the bit which is intended to be written on the next place. On the other hand, the head field must be small enough to avoid re-magnetization of the last written bit , that is due to the fact that when the previous written bit is influenced by the head field of the next bit, its polarity could become reversed and since adjacent magnetic domains could have opposite polarities, the flux reversal might occur while moving from one domain to the next, and may finally result in a few write errors.

The noise in magnetic recording can occurr at any stage of this process. Important factors contributing to noise creation are heads, electronics and media. Magnetic head could make random noise as any other electrically resistive component does. Another source of error is due to the media and the impurities in magnetic material and the variation in the particle size, etc. In addition, during the write process if the magnetic properties of the material are poor, the permanent errors could be seen.
\vspace{-2pt}

To sum up, it seems necessary to consider an extra source of error. Particularly, when the applied head field is smaller than the switching field of the current grains , this bit will be missed and therefore, the prerecorded bit or the last written bit will be recorded in the current position. Hence, this new write channel model will be a generalized version of write channel model in \cite{iyengar2011writechannel}. In a particular case, if we eliminate the effect of this extra binary noise, the model in \cite{iyengar2011writechannel} will be achieved.

\textbf{Notations}: In this paper, all the random variables are shown in capital letters $X, Y, Z, W$, which are defined as channel input, output, state and noise, respectively. We also show random vectors as $Y_i^j = \left(Y_i, Y_{i+1}, ..., Y_{j}\right)$, $i, j \in \mathrm{N}$. The alphabets, on which the random variables are defined, are shown as $\mathbb{X}, \mathbb{Y}, \mathbb{Z}, \mathbb{W}$, and we consider the binary set $\left\{0, 1\right\}$ for the alphabet of all random variables. For simplicity, we define $\bar{p} = 1 - p$, $\bar{\alpha} = 1 - \alpha$ and $\bar{\beta} = 1 - \beta$ as the complements of real numbers $p$, $\alpha$ and $\beta$, respectively; and $h_2\left(.\right)$ denotes the binary entropy function \cite{coverbook}.

\textbf{Paper organization:} The remaining parts of this paper are organized as follows.  The model and main results are explained in section II. And, section III provides concluding remarks.




\section{The Model Statement and Main Results}
\subsection{The Model Statement}
As it was mentioned before, in [11], a write channel model was proposed that reflected two main types of errors including insertion-deletion errors and substitution-like errors. To clarify these two kinds of errors in one closed relation, the rule of writing was introduced by the following expression
\begin{equation}
Y_i = X_{i-Z_i}
\end{equation}
There are three random variables present in this model, input ($X$), output ($Y$), and state ($Z$). It is clear that when $Z_i = 1$ then $Y_i = X_{i-1}$ so that the output stems from the previous input, which might produce an error. In our new model, we add a binary noise to (1), to describe the extra source of error explained in (our work). Therefore, the model is expressed as:
\begin{eqnarray}
Y_i = X_{i-Z_i} \oplus W_i ,
\end{eqnarray}
where $W_i$ shows the binary noise resulting in an extra write error.

At this stage, we consider the Bernoulli state channel and attempt to find bounds for the information rate of channel according to this new proposed model. Here we assume that state and noise random variables have Bernoulli distributions with parameter $p$ and $\alpha$, respectively, and the input $X_i$, the state $Z_i$ and the noise $W_i$ are mutually independent. Thus, we try to find some information-theoretical properties of this channel. Since this channel has memory, we define its capacity as
\begin{eqnarray}
C\left(\alpha, p\right) = \mathop{\lim}_{n \rightarrow \infty} \mathop{\sup}_{p\left\{X_1^n\right\}} \frac{1}{n} I \left(X_1^n; Y_1^n\right).
\end{eqnarray}
The capacity (3) will be derived in terms of $p$ and $\alpha$ as two Bernoulli parameters defining the channel. Consider
\begin{eqnarray}
\frac{1}{n} I\left(X_1^n; Y_1^n\right) &=& \frac{1}{n} H\left(Y_1^n\right) - \frac{1}{n} H\left(Y_1^n \mid X_1^n\right) \nonumber \\
&\overset{\left(\ast \right)}{=}& \frac{1}{n} H\left(Y_1^n\right) - \frac{1}{n} \sum_{i=1}^{n} H\left(Y_i \mid X_{i-1}^{i}\right),
\end{eqnarray}
where $\left(\ast \right)$ is obtained from the rule of writing introduced in (2). By considering the fact that $Y_i$ depends on two random variables $Z_i$ and $W_i$, with Bernoulli distributions as $Z_i \sim B\left(p\right)$ and  $W_i \sim B\left(\alpha\right)$, and by expanding the second term in (4), the following equation is achieved:
\begin{IEEEeqnarray}{llll}
 H \left(Y_i \mid  X_{i-1}^{i}\right) = H \left(Y_i \mid X_{i-1} = X_i = 0\right) \mathrm{P}\left(X_{i-1} = X_i = 0\right) \nonumber \\
\qquad \quad + H \left(Y_i \mid X_{i-1} = X_i = 1\right) \mathrm{P}\left(X_{i-1} = X_i = 1\right) \nonumber \\
\qquad \quad + H \left(Y_i \mid X_{i-1}=0, X_i = 1\right) \mathrm{P}\left(X_{i-1}=0, X_i = 1\right) \nonumber \\
\qquad  \quad + H \left(Y_i \mid X_{i-1}=1, X_i = 0\right) \mathrm{P}\left(X_{i-1}=1, X_i = 0\right)
\end{IEEEeqnarray}
After computing each entropy function in (5), and putting them in (4), the ultimate expression will be derived as follows
\begin{IEEEeqnarray}{lll}
\frac{1}{n} I \left(X_1^n; Y_1^n\right)   &=&   \frac{1}{n} H \left(Y_1^n\right) - \frac{1}{n}
\left[ h_2 \left(\alpha \right) \sum_{i=1}^{n} \mathrm{P} \left\{X_i = X_{i-1}\right\} \right. \nonumber \\
&+&  \left. h_2 \left(p+ \alpha - 2 \alpha p\right) \sum_{i=1}^{n}\mathrm{P}\left\{X_i \neq X_{i-1}\right\rbrace \right].
\end{IEEEeqnarray}
\subsection{An Information Rate Lower Bound}
\subsubsection{i.i.d Input Process}

Lower bounds for information rate can be found by making an assumption about the probability distribution for input. Therefore, at this stage we assume that the input is an i.i.d (independent and identically distributed) process. So due to the binary alphabet of this system, we consider input distribution as a uniform distribution called i.u.d (independent and uniformly distributed). The rate derived with this assumption is known as \emph{symmetric information rate} (SIR).

If we ignore the data dependence of the noise in the previously defined channel we can derive a lower bound for SIR by considering it as a BSC (binary symmetric channel) with error probability as
\begin{IEEEeqnarray}{lll}
\mathrm{P}_{\mathrm{error}} &=& \mathrm{P} \left(Y_i \neq X_i\right) \nonumber \\
&=& \mathrm{P} \left(Z_i = 0, W_i = 1\right) + \mathrm{P} \left(Z_i=1, W_i =0, X_i \neq X_{i-1}\right) \nonumber \\
&+& \mathrm{P} \left(Z_i =1, W_i =1, X_i = X_{i-1}\right) = \frac{p}{2}+ \left(1-p\right) \alpha .
\end{IEEEeqnarray}
Using (7), we are able to find a straight lower bound for SIR as follows
\begin{eqnarray}
C_{iud} \left(\alpha, p\right) \geq 1 - h_2 \left(\frac{p}{2} + \left(1-p\right) \alpha \right) = L_{0}^{iud} \left(\alpha, p\right).
\end{eqnarray}
To find a tighter lower bound, we prove the following lemma.
\begin{lemma}
For the channel model in (2), the following lower bound is obtained
\begin{IEEEeqnarray}{ll}
h_2 \! \left(\! \frac{1-p+2\alpha p}{2}\! \right) \! -\!  \frac{h_2\! \left(\alpha\right)}{2}\! -\!  \frac{h_2 \left(p+\alpha - 2\alpha p\right)}{2} = L_{1}^{iud}\! \left(\alpha, \beta \right)\nonumber \\
\end{IEEEeqnarray}
\end{lemma}
\begin{IEEEproof}
By starting with (6) and conditioning the entropy of output we have
{\small
\begin{IEEEeqnarray}{lll}
C_{iud} \left(\alpha, p \right)  &=& \! \! \mathop{\lim}_{n \rightarrow \infty} \frac{1}{n} \sum_{i=1}^{n}\!  H\!  \left(Y_i \!  \mid \! Y_{1}^{i-1}\right)\!  -\!  \frac{h_2\left(\alpha \right)}{2}\!  - \! \frac{h_2 \left(p+\alpha - 2 \alpha p\right)}{2} \nonumber \\
&\geq &\mathop{\lim}_{n \rightarrow \infty}\! \frac{1}{n}\!  \sum_{i=1}^{n}\! \!  H\!  \left(\! Y_i \!  \mid \!  Y_{1}^{i-1}\! , \! X_{i-1}\! \right) \!\! - \! \! \frac{h_2\! \left(\alpha \right)}{2} \! -\!  \frac{h_2 \! \left(p \! +\!  \alpha \! -\!  2 \alpha p \right)}{2}.\nonumber \\
\end{IEEEeqnarray}}
Now according to (2), we use the independence of $Y_i$ from the other variables except $X_i$ and $X_{i-1}$. So, when $X_{i-1}$ is given, there is no dependency between $Y_i$ and $Y_{1}^{i-1}$. By computing
\begin{IEEEeqnarray}{llll}
& \! \! \! \! \! \! \! \! \! \! \! \! \mathrm{P} \left(Y_i = 0\mid X_{i-1}=0 \right) = \frac{\mathrm{P} \left(Y_i =0, X_{i-1} = 0\right)}{\mathrm{P}\left(X_{i-1} =0\right)} \nonumber \\
& = \frac{1}{0.5} \left[ \mathrm{P}  \left(  W_i = 0 ,  Z_i  = 0,  X_i= 0 , X_{i-1} =0 \right)  \right. \nonumber \\
 &+ \, \mathrm{P} \left(W_i=1, Z_i=0, X_i=1, X_{i-1}=0\right) 
\nonumber \\
& \left. +\, \mathrm{P}\left( W_i = 0, Z_i = 1, X_{i-1}=0 \right) \right]
= \frac{\left(1+p-2 \alpha p\right)}{2}, \nonumber
\end{IEEEeqnarray}
and the other contributing terms of (10), the lemma will be proved.
\end{IEEEproof}

\emph{Corollary} 1: If we put $\alpha = 0$ in (9), the lower bound in [11] is obtained.

\emph{Corollary} 2: As it is evident in Fig. 1, the lower bound shown in (9) is smaller than the corresponding lower bound in [11] with regard to the i.u.d. input. This is because we add an extra noise to this new model so that the lower bound decreases. In addition, as $\alpha$ increases, the difference between these two lower bounds grows.

\subsubsection{First Order Markov Input Process }

As is evident, the channel has memory by nature. So, at this part we assume a memory for input by considering a symmetric first-order binary Markov process $X \sim M_1^{\left(2\right)} \! \left(\beta\right)$ and define it as $\mathrm{P} \left\{X_i = 0 \mid X_{i-1}=1\right\} = \mathrm{P} \left\{X_i=1 \mid X_{i-1} = 0\right\} = \beta$.

\begin{lemma}
As a result, by starting with (6) and considering the first order Markov input, a lower bound for the Symmetric Markov-1 Rate (M1R) is derived as follows:
{\small
\begin{IEEEeqnarray}{lll}
\! C_{M1}\left(\alpha, \beta, p\right) = {H\left(Y\right)} \lvert_{{X \sim M_1^{\left(2\right)}}} -\beta h_2 \left(p+\alpha - 2 \alpha p\right) - \bar{\beta} h_2\left(\alpha\right) \nonumber \\
\, \quad \geq  \mathop{\lim}_{n \rightarrow \infty} \frac{1}{n} \sum_{i=1}^{n} H\!  \left(\! Y_i \mid Y_1^{i-1}, X_{i-1}\! \right) \! -\!  \beta h_2\! \left(p+\alpha \! -\!  2 \alpha p\right)\! - \! \bar{\beta} h_2\!  \left(\alpha
\right)\nonumber \\
\, \quad = h_2 \left(\bar{p} \left(\bar{\alpha} \beta + \alpha \bar{\beta}\right)+ \alpha p\right) - \beta h_2 \left(p+\alpha -2 \alpha p\right)- \bar{\beta} h_2\left(\alpha \right) \nonumber \\
\, \quad  = L_{1}^{M1}\left(\alpha, \beta, p\right).
\end{IEEEeqnarray}}
\end{lemma}
\begin{IEEEproof}
\begin{figure}[t]
\begin{center}$
\begin{array}{cc}
\includegraphics[width=43mm]{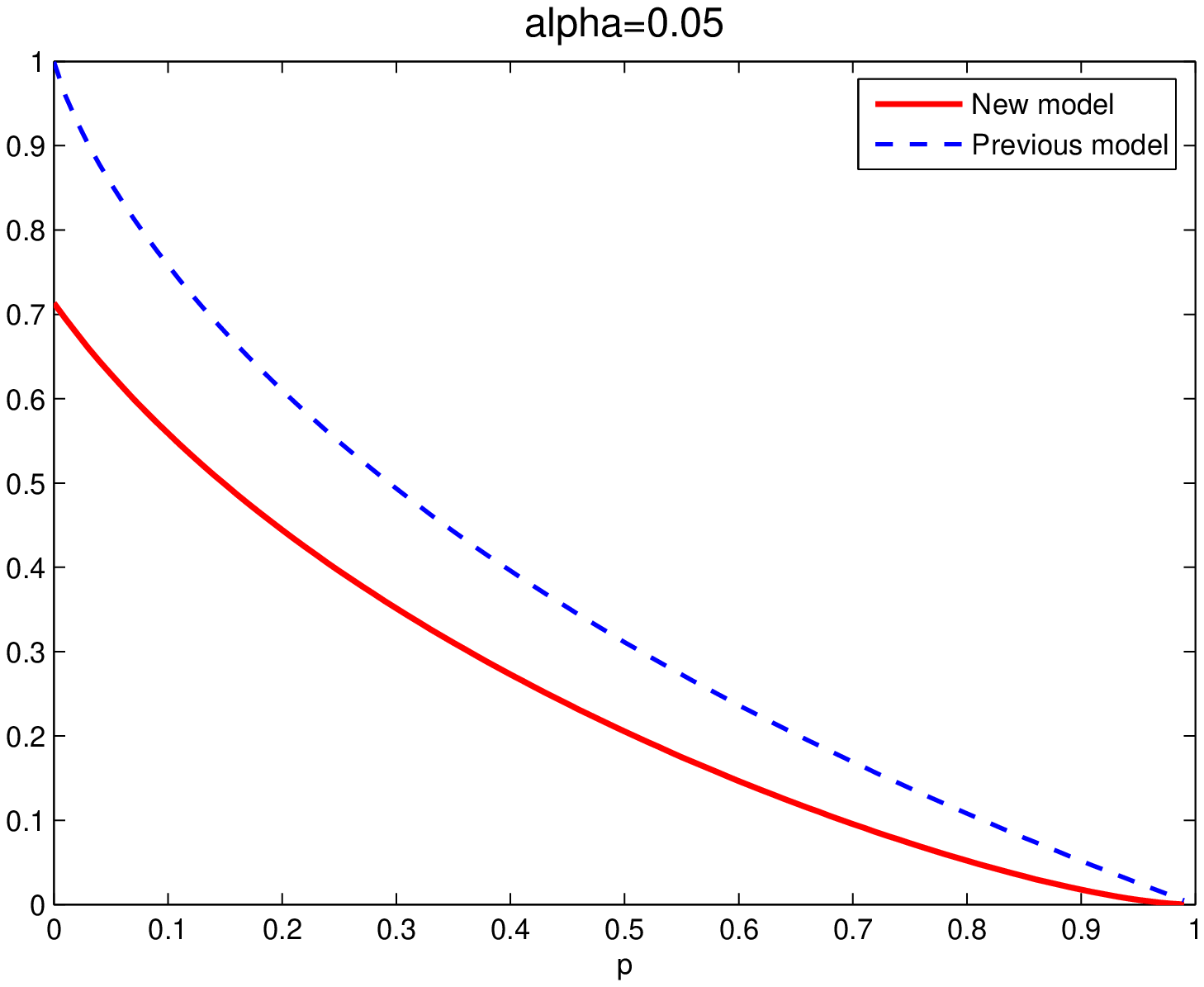}&
\includegraphics[width=43mm]{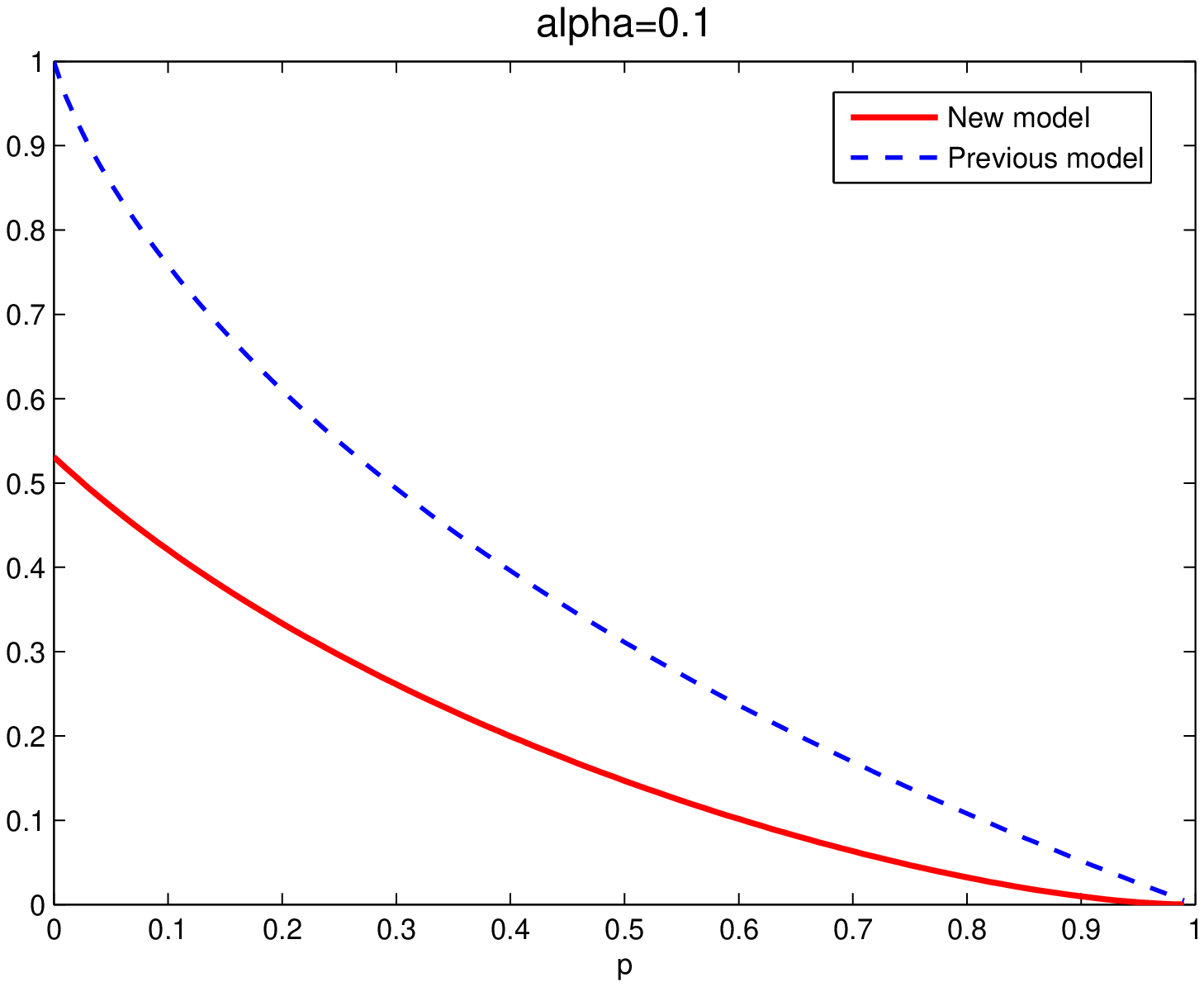} \\
\includegraphics[width=43mm]{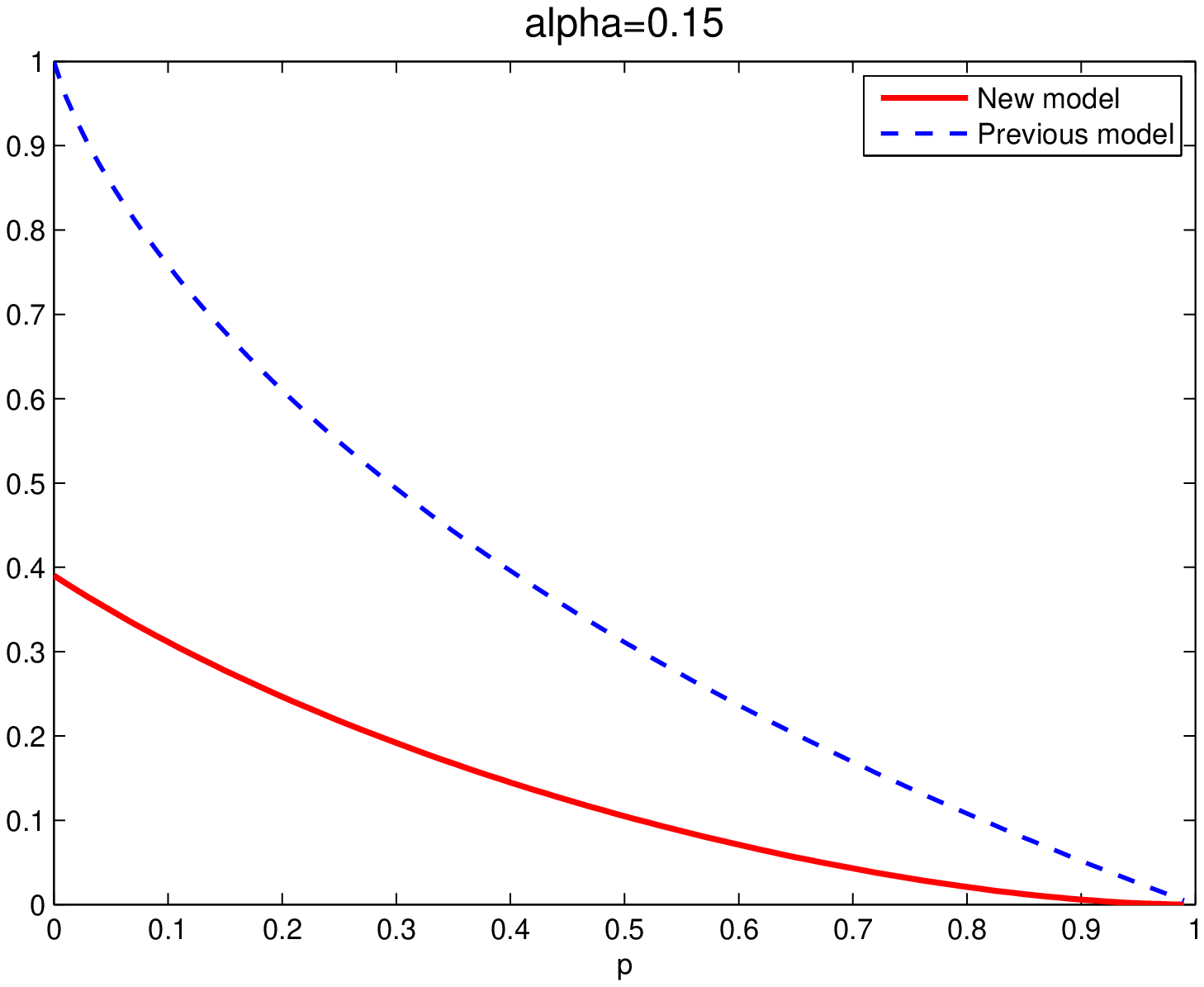}&
\includegraphics[width=43mm]{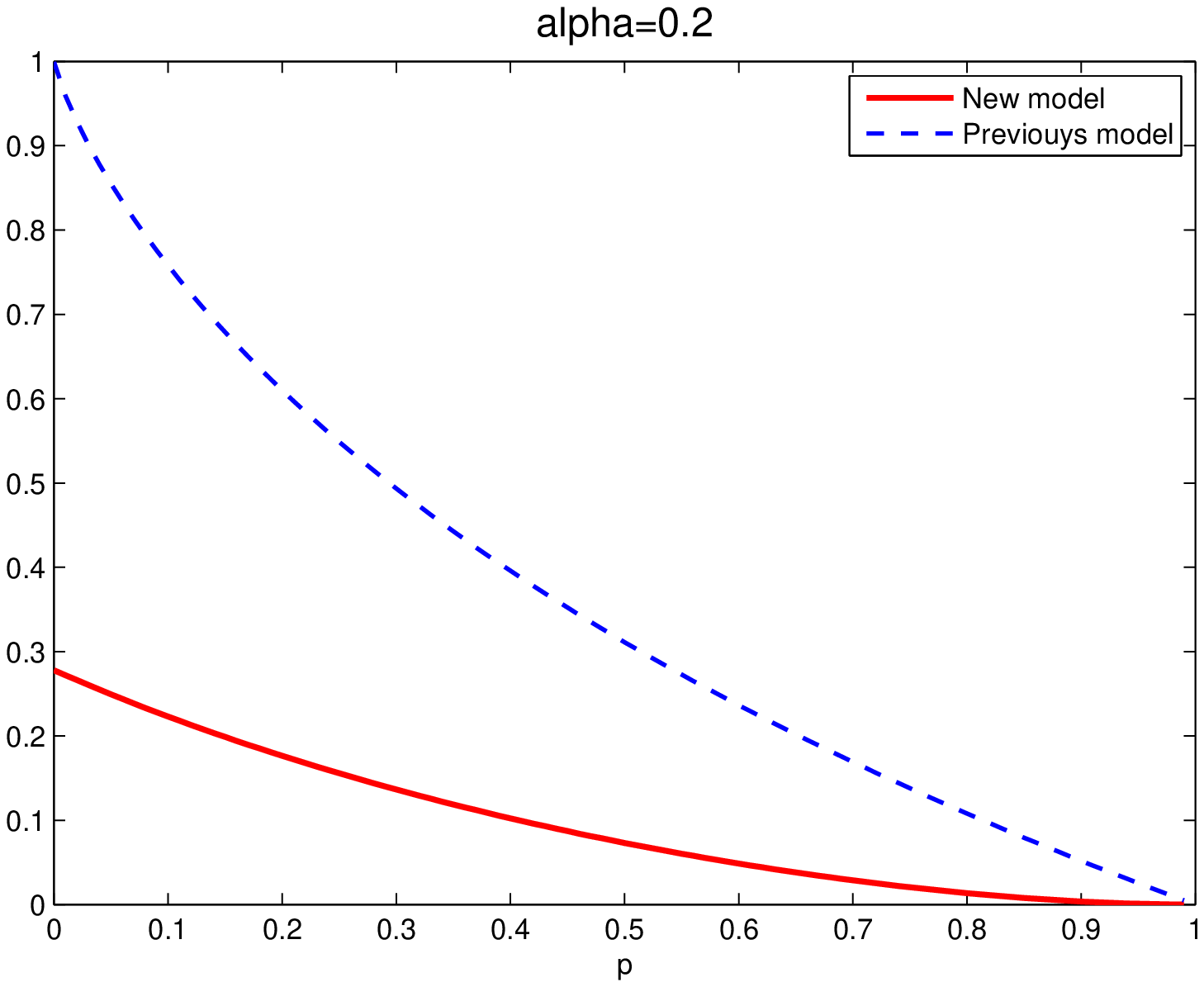}
\end{array}$
\end{center}
\caption{Symmetric information rate lower bounds for the new and last model, with $\alpha = 0.05, 0.1, 0.15, 0.2$. }
\label{Fig1}
\end{figure}
As we mentioned before, there is a type of independence between $Y_i$ and $Y_{1}^{i-1}$ when $X_{i-1}$ is known. So, by starting with (10), we need to compute $H\left(Y_i \mid X_{i-1}=0\right)$. For instance, one of the involved terms in $H\left(Y_i \mid X_{i-1}=0\right)$ is derived as:
{\small
\begin{IEEEeqnarray}{llll}
\mathrm{P} \left(Y_i =1 \mid X_{i-1}=0\right) &=& \mathrm{P}\left(W_i=0, Z_i =0, X_i=1 \mid X_{i-1}=0 \right) \nonumber \\
&+& \mathrm{P} \left(W_i =1, Z_i=0, X_i=0 \mid X_{i-1}=0\right) \nonumber \\
&+&\mathrm{P}\left(W_i=1, Z_i=1,X_{i-1}=0\mid X_{i-1}=0\right)\nonumber \\
&=& \bar{p} \left(\bar{\alpha} \beta + \alpha \bar{\beta}\right)+\alpha p,
\end{IEEEeqnarray}}
so, the proof is completed.
\end{IEEEproof}
\emph{Corollary} 3: By substituting $\alpha =0$ in (11), the lower bound in [11] is obtained.
\subsection{An Information Rate Upper Bound}
\subsubsection{i.i.d Input Process}
By using the fact that the value of entropy function is not higher than unity, we can achieve one of the straight upper bounds for the SIR, implied by (6) as follows
\begin{eqnarray}
C_{iud} \left(\alpha, p\right) \leq 1 \! -\!  \frac{h_2\! \left(\alpha \right)}{2} - \frac{h_2\! \left(p+\alpha -2\alpha p\right)}{2}= U_{0}^{iud}\! \left(\alpha, p\right)
\end{eqnarray}
\begin{figure}[t]
\begin{center}$
\begin{array}{ll}
\includegraphics[width=43mm]{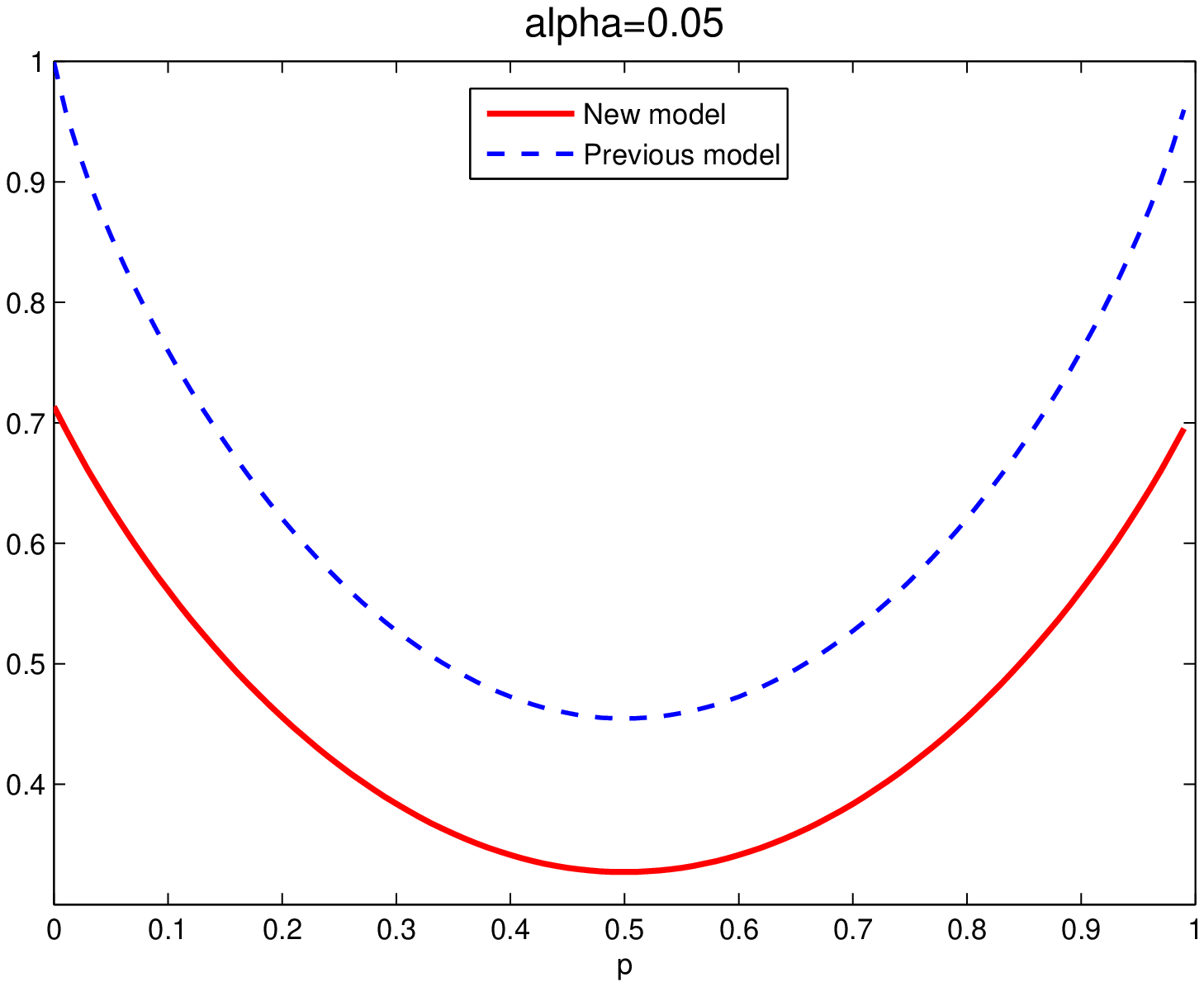}&
\includegraphics[width=43mm]{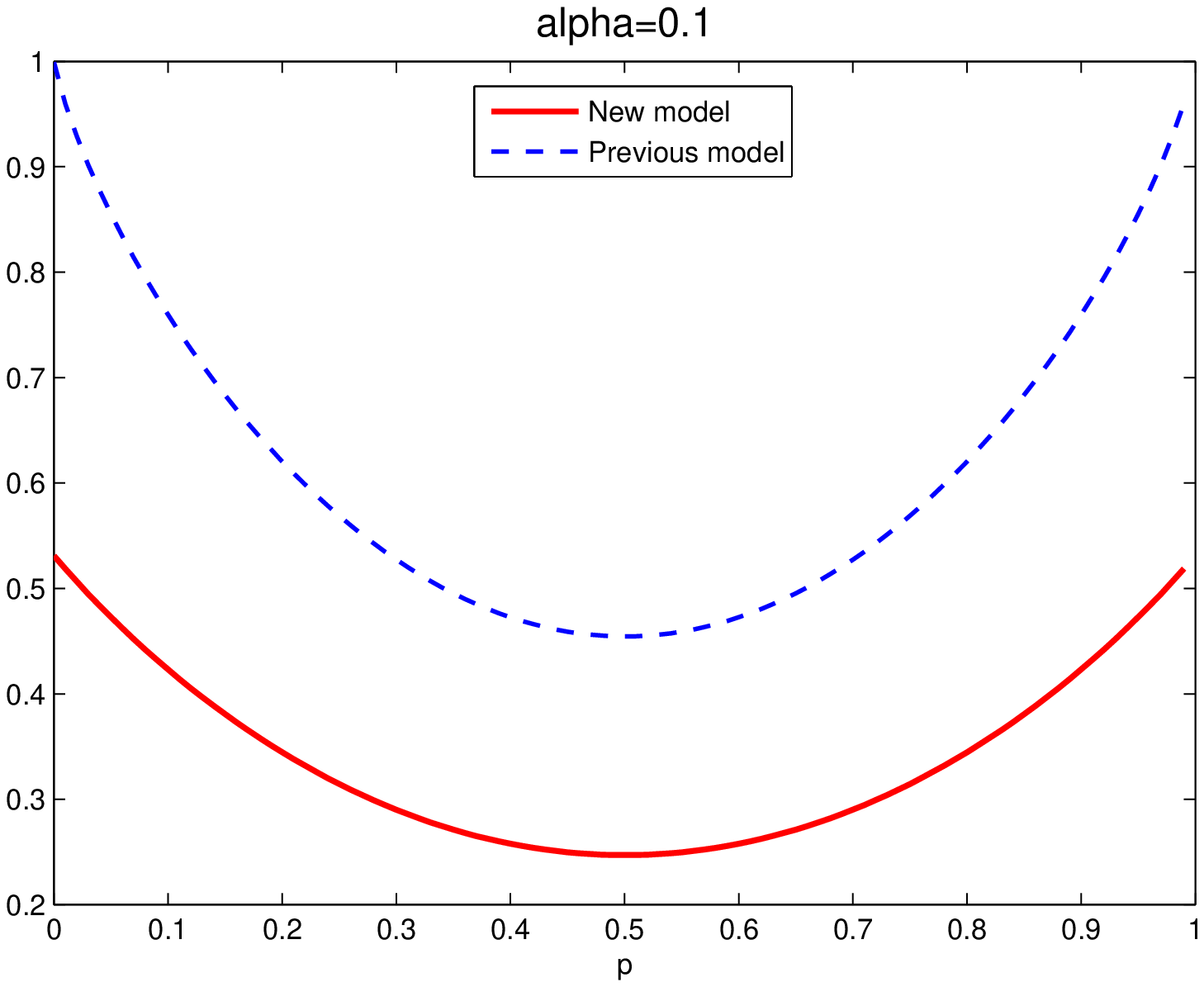} \\
\includegraphics[width=43mm]{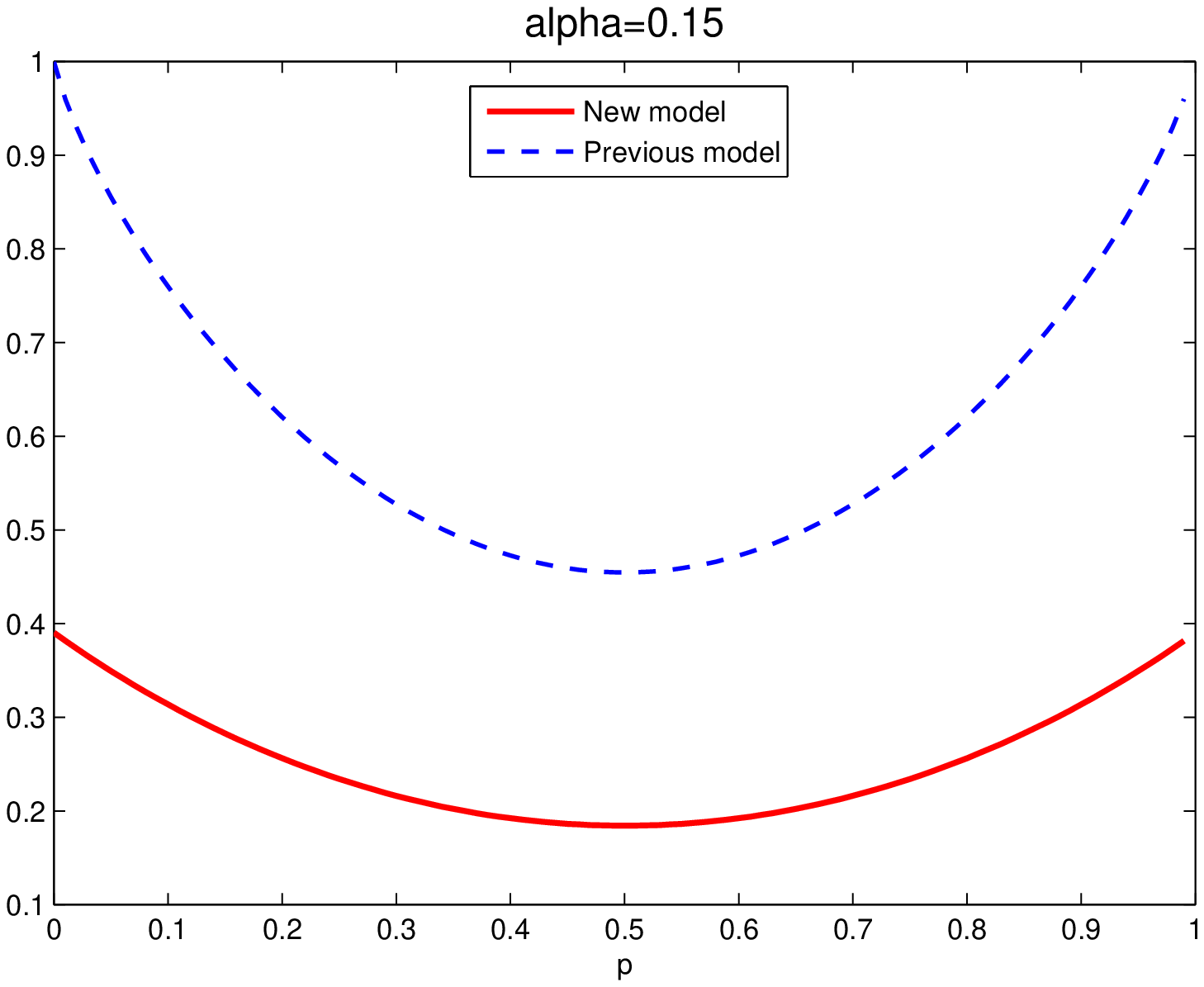}&
\includegraphics[width=43mm]{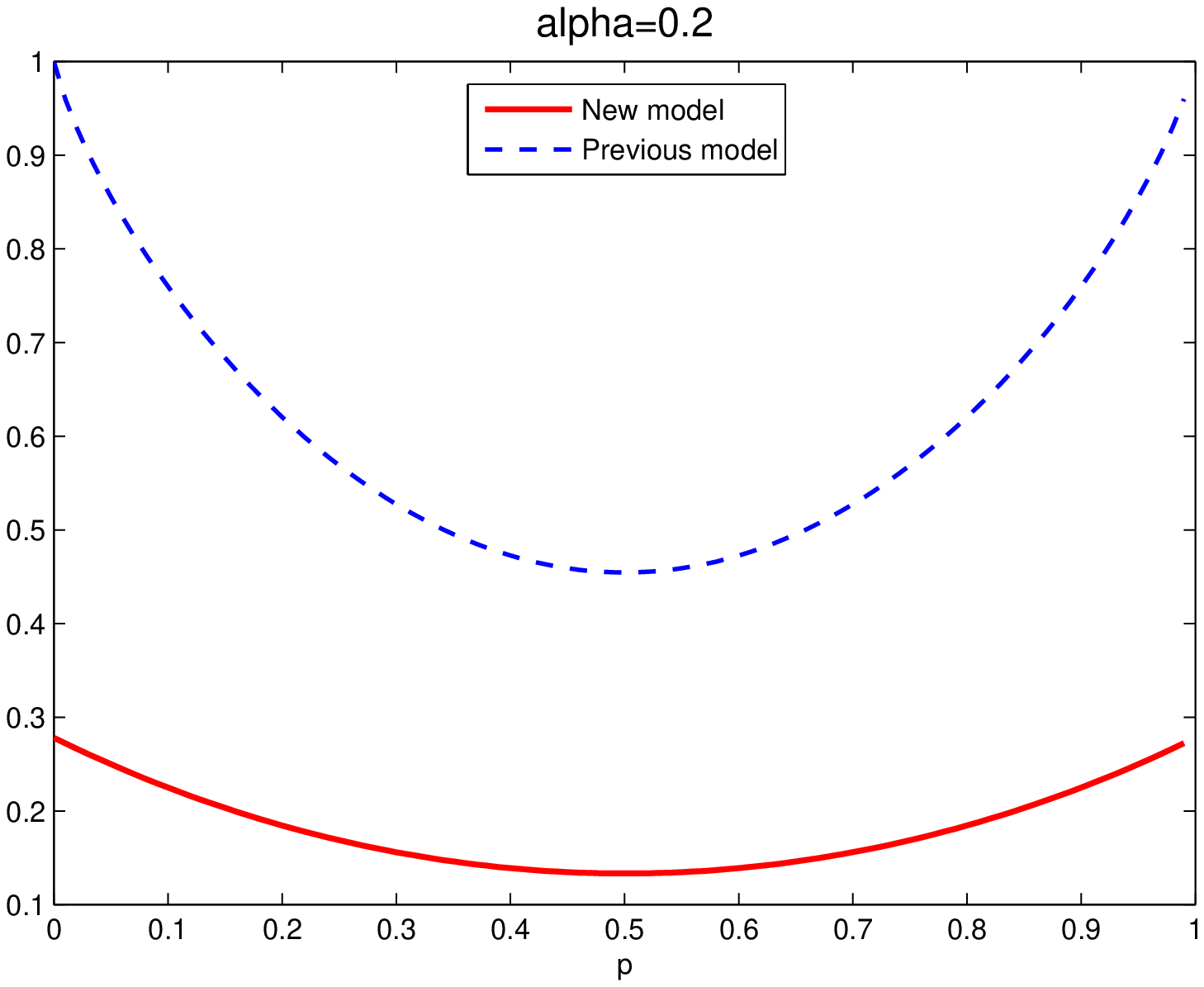}
\end{array}$
\end{center}
\caption{Symmetric information rate upper bounds of the new and last model, with $\alpha = 0.05, 0.1, 0.15, 0.2$. }
\label{Fig2}
\end{figure}
\begin{lemma}
The further upper bound of SIR for channel model in (2) is obtained as follows:
\begin{eqnarray}
h_2\left(\frac{\left(1-p\bar{p}\right)\left(1+2\alpha \left(\alpha - 1\right)\right)+\left(1+p \bar{p}\right) \left(2 \alpha \left(1-\alpha \right)\right)}{2}\right) \nonumber \\
- \frac{h_2\left(\alpha\right)}{2} - \frac{h_2\left(p+\alpha - 2\alpha p\right)}{2} = U_{1}^{iud}\! \left(\alpha, p\right)
\end{eqnarray}
\end{lemma}
\begin{IEEEproof}
Consider
{\small
\begin{eqnarray}
C_{iud}\! \left(\alpha, p\right)\!  \leq \! \! \mathop{\lim}_{n \rightarrow \infty}\! \frac{1}{n} \! \sum_{i=1}^{n}\! H\! \left(\! Y_i \! \mid \!  Y_{i-1}\! \right) \! -\! \frac{h_2\! \left(\alpha\right)}{2} \! -\! \frac{h_2\! \left(p\! +\! \alpha \! -\! 2\alpha p \right)}{2},
\end{eqnarray}}
one of the involved terms in (15) is $\mathrm{P}\left(Y_i =1 \mid Y_{i-1}=1\right)$, therefore, we can expand it as follows:
{\small
\begin{IEEEeqnarray}{llllll}
\mathrm{P}\left(Y_i =1\mid Y_{i-1}=1\right)= \frac{1}{0.5} \nonumber \\
\quad \times [ \mathrm{P} \left(W_i=0, Z_i=0, X_i=1, W_{i-1}=0, Z_{i-1}=0, X_{i-1}=1\right) \nonumber \\
\quad +\mathrm{P}\left(W_i=0, Z_i=0, X_i=1, W_{i-1}=0, Z_{i-1}=1, X_{i-2}=1\right) \nonumber \\
\quad +\mathrm{P}\left(W_i=0, Z_i=0, X_i=1, W_{i-1}=1, Z_{i-1}=0, X_{i-1}=0\right) \nonumber \\
\vspace{4pt}
\quad +\mathrm{P}\left(W_i=0, Z_i=0, X_i=1, W_{i-1}=1, Z_{i-1}=1, X_{i-2}=0\right) \nonumber \\
\quad +\mathrm{P}\left(W_i=0, Z_i=1, W_{i-1}=0, Z_{i-1}=0, X_{i-1}=1\right) \nonumber \\
\quad +\mathrm{P}\left(W_i=0, Z_i=1, X_{i-1}=1, W_{i-1}=0, Z_{i-1}=1, X_{i-2}=1\right) \nonumber \\
\vspace{4pt}
\quad +\mathrm{P}\left(W_i=0, Z_i=1, X_{i-1}=1, W_{i-1}=1, Z_{i-1}=1, X_{i-2}=0\right) \nonumber \\
\quad +\mathrm{P}\left(W_i=1, Z_i=0, X_i=0, W_{i-1}=0, Z_{i-1}=0, X_{i-1}=1\right) \nonumber \\
\quad +\mathrm{P}\left(W_i=1, Z_i=0, X_i=0, W_{i-1}=0, Z_{i-1}=1, X_{i-2}=1\right) \nonumber \\
\quad +\mathrm{P}\left(W_i=1, Z_i=0, X_i=0, W_{i-1}=1, Z_{i-1}=0, X_{i-1}=0\right) \nonumber \\
\vspace{4pt}
\quad +\mathrm{P}\left(W_i=1, Z_i=0, X_i=0, W_{i-1}=1, Z_{i-1}=1, X_{i-2}=0\right) \nonumber \\
\quad +\mathrm{P}\left(W_i=1, Z_i=1, X_{i-1}=0, W_{i-1}=0, Z_{i-1}=1, X_{i-2}=1\right) \nonumber \\
\quad +\mathrm{P}\left(W_i=1, Z_i=1, W_{i-1}=1, Z_{i-1}=0, X_{i-1}=0\right) \nonumber \\
\vspace{4pt}
\quad +\mathrm{P}\left(W_i=1, Z_i=1, X_{i-1}=0, W_{i-1}=1, Z_{i-1}=1, X_{i-2}=0\right) ]\nonumber \\
\quad = \frac{\left(1+p \bar{p}\right)\left(1+2\alpha \left(\alpha -1\right)\right)+\left(1-p \bar{p}\right)\left(2 \alpha \left(1-\alpha \right)\right)}{2},
\end{IEEEeqnarray}}
therefore, the upper bound in (14) is obtained.
\end{IEEEproof}
\emph{Corollary} 4: Choosing  $\alpha = 0$ in (14), the upper bound in [11] is achieved.

\emph{Corollary} 5: As a comparison between these two models in terms of upper bound, Fig. 2 illustrates how the upper bounds vary by changing the values of $\alpha$ in these models. According to Fig. 2, it can be observed that this new model offers the smaller upper bound for SIR. This is mainly because of considering the extra source of error in this channel. The graph also indicates that as $\alpha$ increases, the upper bound diminishes. Thus, in practice the smaller values of $\alpha$ are considered.

\subsubsection{First Order Markov Input Process}

Again, using the fact that entropy is never higher than unity a straight forward upper bound is derived as follows:
\begin{IEEEeqnarray}{l}
C_{M1}\! \left(\alpha, \beta, p\right) \leq 1\! -\!  \beta h_2\! \left(p\! +\! \alpha\!  -\! 2 \alpha p\right)\! -\!  \bar{\beta} h_2 \! \left(\alpha \right) \! = U_{0}^{M1}\! \left(\alpha, \beta, p\right). \nonumber \\
\end{IEEEeqnarray}
\begin{lemma}
The further upper bound is as follows:
\begin{eqnarray}
h_2\left(\left(\bar{\beta}+2{\beta}^2p\bar{p}\right)\left({\bar{\alpha}}^2+{\alpha}^2\right)+2 \alpha \bar{\alpha} \beta \left(p^2 + {\bar{p}}^2 +2 \bar{\beta} p \bar{p}\right) \right) \nonumber \\
- \beta h_2\left(p+ \alpha -2 \alpha p\right) - \bar{\beta} h_2\left(\alpha \right) = U_{1}^{M1}\! \left(\alpha, \beta, p\right).
\end{eqnarray}
\end{lemma}
\begin{IEEEproof}
Consider
{\small
\begin{IEEEeqnarray}{l}
C_{M1} \! \left(\alpha, \beta, p\right)\!  \leq \! \mathop{\lim}_{n \rightarrow \infty}\!  \frac{1}{n} \sum_{i=1}^{n}\!  H\!  \left( Y_i \! \mid \!  Y_{i-1}\right) \! -\!  \beta h_2\! \left(p+\alpha \! -\!  2 \alpha p\right)\! - \! \bar{\beta} h_2\!  \left(\alpha \right), \nonumber
\end{IEEEeqnarray}}
One of the contributing terms in $H \left( Y_i \mid Y_{i-1}\right)$ is derived as follows:
\begin{eqnarray}
\mathrm{P} \left(Y_i = 0 \mid Y_{i-1} =0\right) = \frac{\mathrm{P}\left(Y_i =0, Y_{i-1}=0\right)}{\mathrm{P}\left(Y_{i-1}=0\right)},
\end{eqnarray}
where the nominator of this relation is extended based on the different states on Table I.
\begin{table}[b]
\caption{Various states resulting in $Y_i =0$ and $Y_{i-1}=0$.}
\centering
\begin{tabular}{|c||c|c|c|c||c|c|c|c|}  \hline
{}  &   \multicolumn{4}{c||}{$Y_i =0$}  &    \multicolumn{4}{c|}{$Y_{i-1}=0$}  \\ \hline
State & $W_i$ & $Z_i$ & $X_i$ & $X_{i-1}$ & $W_{i-1}$ & $Z_{i-1}$ & $X_{i-1}$ & $X_{i-2}$\\ \hline \hline
1 & 0 & 0 & 0 & {}  & 0 & 0 & 0 &{}  \\ \hline
2 & 0 & 0 & 0 & {}  & 0 & 1 & {}  & 0\\ \hline
3 & 0 & 0 & 0 &  {} & 1 & 0 & 1 & {} \\ \hline
4 & 0 & 0 & 0 &  {} & 1 & 1 & {}  & 1\\ \hline
5 & 0 & 1 &  {}  & 0 & 0 & 0 & 0 & {} \\ \hline
6 & 0 & 1 & {}   & 0 & 0 & 1 &  {} & 0\\ \hline
7 & 0 & 1 &  {}  & 0 & 1 & 0 & 1 & {} \\ \hline
8 & 0 & 1 &  {}  & 0 & 1 & 1 & {}  & 1\\ \hline
9 & 1 & 0 & 1 & {}  & 0 & 0 & 0 &  {} \\ \hline
10 & 1 & 0 & 1 &  {} & 0 & 1 & {}  & 0\\ \hline
11 & 1 & 0 & 1 & {}  & 1 & 0 & 1 & {}  \\ \hline
12 & 1 & 0 & 1 & {}  & 1 & 1 & {}  & 1\\ \hline
13 & 1 & 1 & {}  & 1 & 0 & 0 & 0 & {} \\ \hline
14 & 1 & 1 & {}  & 1 & 0 & 1 & {}  & 0\\ \hline
15 & 1 & 1 & {}  & 1 & 1 & 0 & 1 & {} \\ \hline
16 & 1 & 1 & {}  & 1 & 1 & 1 & {}  & 1\\ \hline \hline
\end{tabular}\label{tab1}
\end{table}

If we assume an arbitrary value such as 0.5 for the probabilities of assigning initial values (0 or 1) to the input, and consider $X_i$, $Z_i$ and $W_i$ are independent of one another, we can find the value of $\mathrm{P}\left(Y_i = 0\right)$ as follows:
\begin{IEEEeqnarray}{lrr}
\mathrm{P}\!  \left(Y_i=0\right) = \mathrm{P}\! \left(W_i=0,Z_i=0, X_i=0\right)  \nonumber \\
+ \mathrm{P}\! \left(W_i=0,Z_i=1, X_{i-1}=0\right)
+ \mathrm{P}\! \left(W_i=1,Z_i=0, X_i=1\right)  \nonumber \\
+ \mathrm{P}\! \left(W_i=1,Z_i=1, X_{i-1}=1\right) = \frac{1}{2}.
\end{IEEEeqnarray}
Considering the Markov input process, we have
\begin{eqnarray}
\mathrm{P}\! \left(X_i \! =\! 0, X_{i-2}\! =\! 0\right) = \mathrm{P}\!  \left(X_{i-2}\! =\! 0\right) \mathrm{P}\! \left(X_i \! =\! 0\!  \mid \!  X_{i-2}\! =\! 0\right),
\end{eqnarray}
where
\begin{eqnarray}
\mathrm{P}\left(X_i =0 \mid X_{i-2} =0\right) = {\bar{\beta}}^2 + \beta^2.
\end{eqnarray}
By substituting the results of TABLE~\ref{tab1}, (20) and (22) into (19), the lemma will be proved.
\end{IEEEproof}
\emph{Corollary} 6:
Choosing $\alpha =0$ in (18), the upper bound in [11] according to Markov-1 input distribution is achieved.
%
\subsection{The Gap Between Lower and Upper Bounds}

\begin{figure}[t]
    \centering
    \includegraphics[width=3.3in]{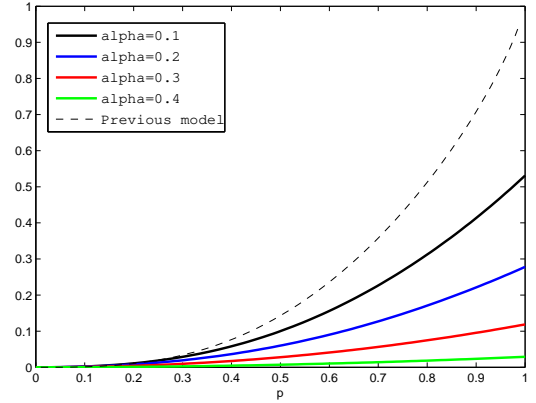}
    \caption{Gap between lower and upper bounds for SIR derived in (9) and (14), and previous write channel model was introduced in [11].}
    \label{Fig4}
\end{figure}

The gap between bounds of SIR in the new and the previous models is introduced as follows:
\begin{IEEEeqnarray}{lll}
\mathrm{Gap}_{\mathrm{new}} &=& U_{1}^{iud}\! \left(\alpha , p\right) - L_1^{iud}\! \left(\alpha , p\right) \nonumber \\
&=& h_2\! \left(\!\! \frac{\left(1-p\bar{p}\right)\left(1\! -\! 2\alpha \left(\alpha -1\right)\right)\! +\! \left(1+p\bar{p}\right)\left(2\alpha \left(1- \alpha\right)\right)}{2}\! \! \right)\nonumber  \\&-& h_2\! \left(\frac{1-p+2 \alpha p}{2}\right),\\
\mathrm{Gap}_{\mathrm{old}} &=& h_2\left(\frac{1-p\bar{p}}{2}\right)-h_2\left(\frac{1-p}{2}\right).
\end{IEEEeqnarray}
Since the capacity for such channels is still unknown, it is worth reaching to a smaller gap between the lower and upper bounds. As it is evident in Fig.~3, gap is widening as $p$ increases, this is due to the fact that a larger value of $p$ results in a higher probability of error, because, this parameter presents the probability of dependency of each bit on its previous written bit, which can act as a source of error in such a channel. To explore the advantages that this new channel model offers, we can compare the derived gap in (23) with the gap shown in (24). Fig.~3 illustrates the fact that in spite of the smaller lower bound, this new write channel model is closer to the SIR, due to the smaller gap. By using this new write channel model we are able to estimate an expression for SIR in terms of lower and upper bounds with lower probability of error.

\section{Conclusion}
We proposed a new write channel model, which consists of both media deflections that cause some random errors, and insertion-deletion, substitution-like errors that come from the necessity of synchronization. It is evident that this channel is a generalized version for the model proposed in [11]. As it was mentioned before, $\alpha$ determines the probability of assigning $1$ to $W_i$, so if $\alpha$ grows to reach $0.5$, it would be expected to encounter a low efficiency channel. Therefore, in practice it is justified to consider small values for $\alpha$, in order to achieve the qualified channel; however, from Fig.~3 it is evident that this channel can offer the smaller gap even for small values of $\alpha$.

We also derived some information-theoretic properties for this Bernoulli state channel according to the i.u.d. and Markov-1 input process and proved that this new model results are the generalization of the previous model and ultimately, for the i.u.d input process, it was shown that although we encounter the smaller lower bound, which is not desired to come close to the SIR, this model can bring smaller upper bound in order to approach the SIR since the gap between these bounds are smaller in this new model, compared with the previous write channel model. There is a fact that the source of the phase mismatch between head field and BPMR dot location may come from both mechanical and process variations where the former has been discussed to be like a correlated process, meaning that once it happens, there will be a higher chance for it to happen in the subsequent dot stream. Thus, in our future work, we plan to consider Markov state channel model and analyze the information rate in order to achieve the desirable results in solving real problems.


%
%



\bibliographystyle{IEEEtran}
\bibliography{refrence}




\end{document}